\documentclass[10pt]{IEEEtran}
\usepackage{epstopdf}
\usepackage{graphicx}
\usepackage{stmaryrd}
\usepackage{amssymb,amsmath}
\usepackage{multirow,url}
\usepackage{color,cite}
\usepackage{hyperref}

\ifCLASSINFOpdf
\else
\fi
\begin{document}
%
\title{Channel-coded Collision Resolution by Exploiting Symbol Misalignment}


%

\author{Lu Lu$^{\ast}$, Soung Chang Liew$^{\ast}$, Shengli Zhang$^{\ast}$$^{\dagger}$\\$^{\ast}$
Department of Information Engineering, The Chinese University of Hong Kong, Hong Kong\\
$^{\dagger}$Department of Communication Engineering, Shenzhen University, China\\
email:\{ll007, soung, slzhang\}@ie.cuhk.edu.hk

\thanks{This work was supported by the Competitive Earmarked Research Grant (Project ID 414507)
established under the University Grant Committee of the Hong Kong
Special Administrative Region, China, and the Direct Grant (Project
ID 2050436) of the Chinese University of Hong Kong. This work is
also supported by National Science Foundation of China (Grant No.
60902016).}}


\maketitle

\begin{abstract}


In random-access networks, such as the IEEE 802.11 network,
different users may transmit their packets simultaneously, resulting
in packet collisions. Traditionally, the collided packets are simply
discarded. To improve performance, advanced signal processing
techniques can be applied to extract the individual packets from the
collided signals. Prior work of ours has shown that the symbol
misalignment among the collided packets can be exploited to improve
the likelihood of successfully extracting the individual packets.
However, the failure rate is still unacceptably high. This paper
investigates how channel coding can be used to reduce the failure
rate. We propose and investigate a decoding scheme that incorporates
the exploitation of the aforementioned symbol misalignment into the
channel decoding process. This is a fine-grained integration at the
symbol level. In particular, collision resolution and channel
decoding are applied in an integrated manner. Simulation results
indicate that our method outperforms other schemes, including the
straightforward method in which collision resolution and channel
coding are applied separately.

\end{abstract}


%
\IEEEpeerreviewmaketitle

\section{Introduction}
%
%

An important issue in wireless random-access networks is how to deal
with the collisions caused by simultaneous packet transmissions.
Multiuser detection (MUD), first proposed by Verdu in the 1980s, is
the traditional method to solve this problem. However, this method
relies heavily on the crosscorrelation of the signature waves, thus
it is mainly used in CDMA channels.

In this paper, we are interested in resolving the collisions in
802.11-like wireless local area network (WLAN). By collision
resolution, we mean extracting the individual signals within an
overlapped signals in a collision. In carrier-sense multiple-access
(CSMA) WLAN, the carrier-sensing mechanism tries to avoid
simultaneous transmissions by different users. However, collisions
can still happen when the backoff process of two or more stations
count down to zero at the same time. Collisions can also happen due
to the hidden-node problem \cite{Hidden Ng, Hidden Jiang} wherein
two stations that cannot carrier-sense each other transmit
simultaneously.

{\it The signature waves of different users are the same in a WLAN},
making collision resolution particular challenging. However, in a
WLAN, when the signals of simultaneously transmitting stations reach
a receiver, their symbols will most likely not be aligned. This
introduces a certain degree of orthogality between the same
signature waveforms. In particular, this symbol misalignment can be
exploited for the extraction of individual signals. In \cite{CRESM}
we proposed an algorithm named CRESM (Collision Resolution by
exploiting Symbol Misalignment) to do so. CRESM has the same
performance as the traditional Asynchronous MUD (A-MUD)
\cite{VerduTrans86} when the signature waveforms of users are the
same, but with smaller complexity.

Although CRESM and A-MUD can minimize the probability of error
optimally \cite{Verdu Trans 86, Verdubook}, the BER is still
unacceptably high. In this paper, we investigate the use of channel
coding on top of CRESM to reduce the error probability.
Channel-coded MUD applied on different signature waves can be traced
back to a decade ago, when Wang and Poor \emph{et al}. \cite{Wang
Poor JASC99} combined the idea of Turbo coding with MUD. This method
is mainly targeted for the CDMA system. We shall refer to this
method as Turbo-SIC (Turbo Soft Interference Cancellation).

An insight from CRESM \cite{CRESM} and traditional MUD \cite{Verdu
Trans 86} is that joint decoding of the symbols may yield better
results in the non-channel-coded case. An outstanding issue is
whether joint decoding will also yield better results in the
channel-coded case. In particular, can we integrate channel decoding
and the exploitation of the symbol misalignment under one framework?

In this paper we present a scheme, named C-CRESM (Channel-coded
CRESM), for this purpose. To illustrate our method, we focus on the
use of the Repeat Accumulate (RA) channel code \cite{RA code Jin
Phd}. By constructing a \emph{virtual} Tanner graph associated with
the misaligned collided bit streams, we derive the message update
rules in the associate belief  propagation algorithm to extract the
collided signals.  Simulation results show that C-CRESM can perform
better than Turbo-SIC. The improvement is particularly significant
when the target BER is below $10^{-3}$.

The remainder of this paper is organized as follows: Section II
gives the system model. Section III presents C-CRESM.  Section IV
compares C-CRESM with other channel-coded MUD methods. Section V
gives simulation results that demonstrate the superiority of C-CRESM
over these other schemes. Section VI concludes this paper.

\section{System Model}

We consider the asynchronous multiple access channel with two end
nodes (node \emph{A} and node \emph{B}) transmitting to one Access
Point (AP), as shown in Fig. 1. Both end nodes employ BPSK
modulation with the RA channel code, and the channel is assumed to
be AWGN. Throughout this paper, we use capital letters to represent
a packet. Specifically, $S_i$, $X_i$ and $Y_i$ denote the uncoded
source packet, channel-coded packet, and received packet of node
$i$, respectively. The lowercase letters, $s_j\in \{ 0,1\}$, $x_k\in
\{ 0,1\}$, and $y_k$ are the corresponding symbols in the packets,
where $1 \le j \le N$ and $1 \le k \le qN$ ($N$ is the total length
of the source packet and \emph{q} is the repeat factor of RA code).

We represent a wireless packet by a stream of discrete complex
numbers \cite{DigitalCommSklar}. Specifically, we denote by complex
numbers $x_A[m]$ and $x_B[m]$ the channel coded symbols of nodes
\emph{A} and \emph{B}, respectively. The symbol duration is
normalized to 1. The relative symbol misalignment between the
collided signals is $\Delta$, where $0 < \Delta  < 1$. The
overlapped signals received at the AP can be expressed as follows
\begin{align}r(t) &= h_A (t)\left( {1 - 2x_A [\left\lfloor t \right\rfloor ]}
\right)\cos (\omega _c t) + \nonumber \\ & h_B (t)\left( {1 - 2x_B
[\left\lfloor {t - \Delta } \right\rfloor ]} \right)\cos (\omega _c
(t - \Delta )) + w(t), \label{eq:systemoverlapped}\end{align} where
$w(t)$ is additive white Gaussian noise with power spectral density
$S_w (f) = N_0 /2$; $h_A(t)$ and $h_B(t)$ are complex numbers
representing the channel gains from \emph{A} and \emph{B} to the AP,
respectively; $\left\lfloor t \right\rfloor $ is the largest integer
no larger than $t$; $\omega_c$ is the carrier angular frequency.

We assume the channel is slow fading so that $h_A(t)$ and $h_B(t)$
stay constant within a packet duration, \emph{i.e.}, $h_A(t)= h_A$
and $h_B(t)= h_B$. To ease exposition, we further assume that the
transmit powers and the channel gains of the two nodes are the same,
\emph{i.e.}, $h_A=h_B=1$. We also assume perfect carrier phase
synchronization. Note that this assumption is not absolutely needed,
and we only use it to simplify the model for easier explanation.
Simulation result in \cite{CRESM} shows that without carrier phase
synchronization we can still decode the collision, albeit with a
small SNR penalty.

\begin{figure}[tt]
\centering
\includegraphics[width=0.35\textwidth]{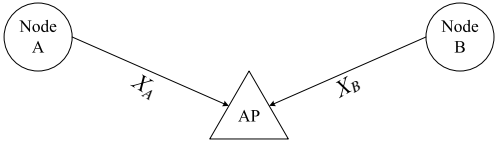}
\caption{System model for two packets collisions} \label{fig:system}
\end{figure}


We use the same digitalization method as in Fig. 3 of \cite{CRESM}
we perform the matched filtering on the overlapped packet in every
$\Delta$ and $1-\Delta$ periods alternatively, with a normalization
factor $1/\Delta$ and $1/(1-\Delta)$ respectively. After
oversampling the received signal we get two parts: the odd part,
which is sampled over a duration of $\Delta$ within a symbol, and
the even part, which is sampled over the rest of the $1-\Delta$
within a symbol. The sampled signals are as follows:
\begin{align}
&r[2k - 1] = (1 - 2x_A [k]) + (1 - 2x_B [k - 1]) + w_{odd} [k]\nonumber \\
&r[2k] = (1 - 2x_A [k]) + (1 - 2x_B [k]) + w_{even} [k],
\label{eq:oversample}\end{align} where $k \ge 1$, $1 - 2x_B [0] =
0$, $w_{odd} [k]$ and $w_{even} [k]$ is a zero mean Gaussian noise
with variance $\Delta N_0 /2$ and $(1-\Delta) N_0 /2$ respectively.

The above set-up is without channel coding. To incorporate channel
coding, we consider the use of the RA channel code . We also focus
on two-packet collisions in this paper. C-CRESM can be easily
extended to deal with collisions with more than two packets.

\section{C-CRESM}
In this section, we construct C-CRESM based on RA (repeat
accumulate) code. This section is organized as follows: First we
give a brief review of the RA code with its decoding algorithm.
Interested readers who wish to learn more about RA code are referred
to \cite{RAconf} for details. Then we construct a \emph{virtual}
Tanner graph, that reveals our C-CRESM can integrate channel coding
with collision resolution. Finally, the message update rules of the
associate belief propagation algorithm are derived to compute the
source information of both nodes, based on the \emph{virtual} Tanner
graph.

\subsection{Review of RA code}
\begin{figure}[tt]
\centering
\includegraphics[width=0.35\textwidth]{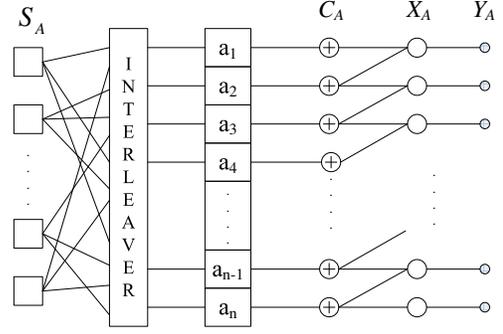}
\caption{Tanner graph for standard RA code} \label{fig:TannerRA}
\end{figure}

The RA code can be decoded by applying the Belief Propagation (BP)
algorithm \cite{RAcodeJinPhd}. The encoding (decoding) Tanner graph
\cite{BP} is shown in Fig. \ref{fig:TannerRA} for a standard RA
code. Generally speaking, the encoding process is like this (from
left to right in Fig. \ref{fig:TannerRA}): the source information
$S_A$ is first repeated $q$ times ($q=3$ in \ref{fig:TannerRA});
then an interleaver is applied to de-correlate the adjacent repeated
bits; finally the punctured bits are accumulated by an \emph{XOR}
operation (represented by nodes $C_A$) sequentially to get the
channel coded bits (nodes $X_A$) for transmission.

The decoding process applies BP on the Tanner graph. The
observations $Y_A$ are given. Based on that, first the channel coded
bits $X_A$ are estimated. Message update rules of BP are then used
to estimate the source information $S_A$.The source information
estimated in this first round are then used to refine the estimation
$X_A$. The right-to-left and left-to-right estimations are iterated
until the estimated $S_A$ converges.

\subsection{Virtual Tanner Graph for RA coded CRESM}
As with CRESM \cite{CRESM}, we construct a \emph{virtual} Tanner
graph for resolving two-user channel-coded collisions under symbol
misalignment. The difference with CRESM is that the transmitted data
are channel-coded here but not in CRESM.

\begin{figure}[tt]
\centering
\includegraphics[width=0.45\textwidth]{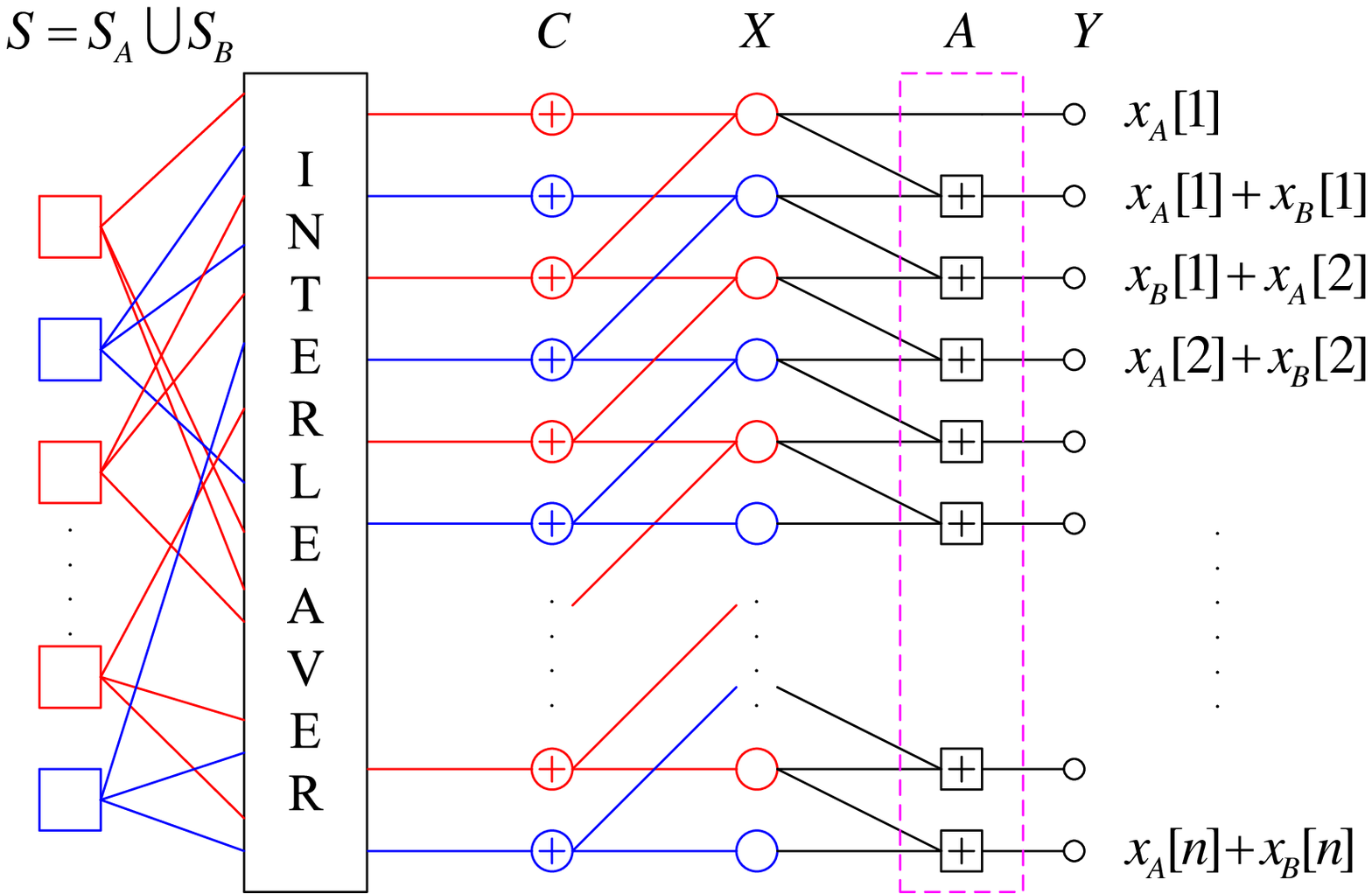}
\caption{\emph{Virtual} Tanner graph for two-user CRESM with RA
code, with the red and blue parts representing the signals from
nodes \emph{A} and \emph{B}, respectively.} \label{fig:TannerCCRESM}
\end{figure}

The collided symbols, after the application of the same oversampling
and digitalization method as in \cite{CRESM}, are $x_A [1],x_A [1] +
x_B [1],x_B [1] + x_A [2], \cdot  \cdot  \cdot \cdot  \cdot ,x_A [n]
+ x_B [n]$, where $x_A[i]$ and $x_B[i]$ represent the repeated
channel coded bits of the two end nodes. This representation ignores
the noise terms that will be added back in our later discussion of
the decoding process. The oversampled signals can be considered as
the outputs of the \emph{virtual} Tanner graph shown in Fig.
\ref{fig:TannerCCRESM}. Note that in Fig. \ref{fig:TannerCCRESM},
the ``square'' add nodes (nodes \emph{A}) are introduced to model
the addition of collided signals when they arrive at the receiver
simultaneously. There are actually no add nodes in the physical RA
coders at the two transmitters. But together, the two RA coders
appear to the receiver as a virtual coder that implements the code
design corresponding to the \emph{virtual} Tanner graph.

Based on the \emph{virtual} Tanner graph in Fig.
\ref{fig:TannerCCRESM}, we now discuss the decoding algorithm. It
should be noted that although we assume BPSK modulation for
simplicity, this algorithm can be generalized to other modulation
schemes. We rewrite the oversampled signals in (\ref{eq:oversample})
as follows:
\begin{align}
&r[2k - 1] = \tilde y[2k - 1] + w_{odd} [k] = 2 - 2y[2k - 1] + w_{odd} [k] \nonumber \\
&r[2k] = \tilde y[2k] + w_{even} [k] = 2 - 2y[2k] + w_{even} [k],
\label{eq:ReceivedDigital}\end{align} where $\tilde y[i]$ is the
\emph{i}-th over-sampled symbol received minus the noise, and $y[i]$
is the unmodulated mixture of signals from nodes A and B that
corresponds to the \emph{i}-th evidence node in the \emph{virtual}
Tanner graph.

\subsection {Definitions}
Before discussing the decoding algorithm, let us first define some
notations and terms. Define nodes \emph{S}, \emph{C}, \emph{X},
\emph{A} and \emph{Y} to be the sets of source nodes, check nodes,
code nodes, add nodes and evidence nodes, respectively. We refer to
the line connecting any two nodes as an `edge'. Define $P_k$ to be
the message from the \emph{k}-th evidence node (a node in \emph{Y})
to the \emph{k}-th add node (a node in \emph{A}), where $1 \le k \le
2n$ and $n=qN$. We represent a message on a non-rightmost edge
(\emph{i.e.}, an edge that is not between a node in \emph{A} and a
node in \emph{Y}) by $Q_k$, $R_k$, or $W_k$.

$P_k  = (p_0 ,p_1 ,p_2 )$ is a probability vector, where $p_i  = \Pr
(y[k] = i{\rm{ }}|{\rm{ }}r[k])$. $Q_k  = (q_1 ,q_2 )$, $R_k  = (r_1
,r_2 )$ and $W_k  = (w_1 ,w_2 )$ are also probability vectors, where
$q_i$, $r_i$ and $w_i$ are $\Pr (x[k] = i)$.  Note that $x[k] = x_A
[{{(k + 1)} \mathord{\left/
 {\vphantom {{(k + 1)} 2}} \right.
 \kern-\nulldelimiterspace} 2}]$ if \emph{k} is odd,
and $x[k] = x_B [k/2]$ if \emph{k} is even.

\subsection {Message Update Rules}
All messages associated with the non-rightmost edges in Fig.
\ref{fig:TannerCCRESM} are initially set to (1/2, 1/2). $P_k  = (p_0
,p_1 ,p_2 )$ is computed based on the received signals $r[k]$ and
will not change throughout the iteration. $P_{2k-1}$ and $P_{2k}$
for $k = 1,...,n$ are given as follows:
\begin{align}
&P_{2k - 1}  = (p_0 ,p_1 ,p_2 ) \nonumber\\
&\begin{aligned} = &(\Pr (y[2k - 1] = 0|r[2k - 1]), \Pr (y[2k - 1] =
1|r[2k - 1]) , \\  &\Pr (y[2k - 1] = 2|r[2k - 1]) ) \\
= &\frac{1}{{\beta [2k - 1]}}\left( \exp ( - \frac{{(r[2k - 1] -
2)^2 }}{{2\sigma ^2 \Delta }}), 2\exp ( - \frac{{(r[2k - 1])^2
}}{{2\sigma ^2 \Delta }}), \right. \\ &\left. \exp ( - \frac{{(r[2k
- 1] + 2)^2 }}{{2\sigma ^2 \Delta }}) \right),
\end{aligned}
\label{eq:initialequ1}\end{align} where $\beta [2k - 1]$ is a
normalized factor given by $\beta [2k - 1] = \exp ( - \frac{{(r[2k -
1])^2 }}{{2\sigma ^2 \Delta }})\left( {\exp (\frac{{2r[2k - 1] -
2}}{{\sigma ^2 \Delta }}) + 2 + \exp ( - \frac{{2r[2k - 1] +
2}}{{\sigma ^2 \Delta }})} \right)$.
\begin{align}
&P_{2k}  = (p_0 ,p_1 ,p_2 ) \nonumber\\
&\begin{aligned} = &(\Pr (y[2k] = 0|r[2k]), \Pr (y[2k] =
1|r[2k]) , \\  &\Pr (y[2k] = 2|r[2k]) ) \\
= &\frac{1}{{\beta [2k]}}\left( \exp ( - \frac{{(r[2k] - 2)^2
}}{{2\sigma ^2 (1-\Delta) }}), 2\exp ( - \frac{{(r[2k])^2
}}{{2\sigma ^2 (1-\Delta) }}), \right. \\ &\left. \exp ( -
\frac{{(r[2k] + 2)^2 }}{{2\sigma ^2 (1-\Delta) }}) \right),
\end{aligned}
\label{eq:initialequ2}\end{align} where $\beta [2k]$ is a normalized
factor given by $\beta [2k] = \exp ( - \frac{{(r[2k])^2 }}{{2\sigma
^2 (1-\Delta) }})\left( {\exp (\frac{{2r[2k] - 2}}{{\sigma ^2
(1-\Delta) }}) + 2 + \exp ( - \frac{{2r[2k] + 2}}{{\sigma ^2
(1-\Delta) }})} \right)$.

Without loss of generality, we omit the time index \emph{k} for
simplicity in the following discussion of message-update rules. We
follow the principles and assumptions of the BP algorithm to derive
the update equations, \emph{i.e.}, the output of a node should be
consistent with the inputs while adopting a 'sum of product' format
of the possible input combinations \cite{UnderstandingBP}.

Based on the above $P_k$ and the initial values of all other
messages, the right-to-left messages in the Tanner graph are first
updated (\ref{31} to \ref{33} below). After that, the left-to-right
messages are updated (\ref{34} to \ref{36} below). The above updates
are iterated until the message values converge. The procedure is
similar to that in \cite{ZhangJASC, RAcodeJinPhd} except for the
modifications required to deal with the virtual add nodes \emph{A}
in Fig. \ref{fig:TannerCCRESM}.

\subsubsection{From add nodes to code nodes}\label{31}
\begin{figure}[tt]
\centering
\includegraphics[width=0.5\textwidth]{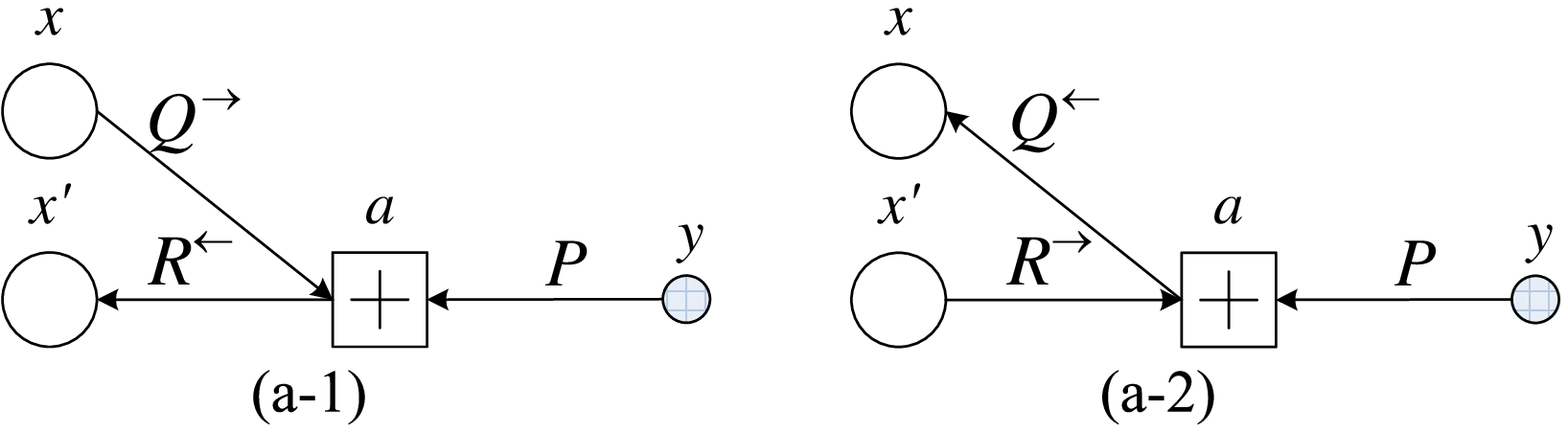}
\caption{\emph{Virtual} Tanner graph for two-user CRESM with RA
code, with the red and blue parts representing the signals from
nodes A and B, respectively.} \label{fig:AddToCode}
\end{figure}

With reference to Fig. \ref{fig:AddToCode}, we derive the update
equations for a right-to-left message $R^ \leftarrow   = (r_0 ,r_1
)$ from an add node to a code node. The update is based on the
existing value of the left-to-right message $Q^ \to   = (q_0 ,q_1 )$
and the fixed message $P = (p_0 ,p_1 ,p_2 )$ from the evidence node.

The symbol $x'$ can only be 0 if $(y,x)=(0,0)$ or $(y,x)=(1,1)$
(\emph{i.e.}, for compatibility, $x+x'$ must be $y$). Based on the
sum-product principle of the BP algorithm, we have
\begin{equation}r_0 = p_0 q_0 + p_1 q_1. \label{eq:addequ1}\end{equation}
Similarly, we have \begin{equation}r_1  = p_2 q_0 + p_2 q_1.
\label{eq:addequ2}\end{equation} Thus, the overall message update
associated with the compatibility requirement of an add node is
given by\begin{equation}R^ \leftarrow = ADD(P,Q^ \to  ) = (p_0 q_0 +
p_1 q_1 ,p_1 q_0  + p_2 q_1 ). \label{eq:addequ3}\end{equation}
Similarly, with reference to Fig. \ref{fig:AddToCode}(a-2), we can
obtain the update equation for message $Q^\leftarrow$ in the
direction of $a$ to $x$:
\begin{equation}Q^ \leftarrow   = ADD(P,R^
\to ) = (p_0 r_0  + p_1 r_1 ,p_1 r_0  + p_2 r_1 ).
\label{eq:addequ4}\end{equation}

\begin{figure}[tt]
\centering
\includegraphics[width=0.5\textwidth]{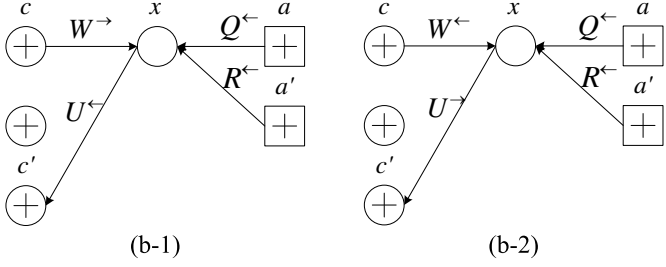}
\caption{\emph{Virtual} Tanner graph for two-user CRESM with RA
code, with the red and blue parts representing the signals from
nodes A and B, respectively.} \label{fig:CodeToCheck}
\end{figure}

Note that in BP, the two messages in opposing directions on the same
edge are distinct. In particular, the input values of $q_i$ in
(\ref{eq:addequ3})and $r_i$ in (\ref{eq:addequ4}) are the values
obtained from the prior iteration for the messages in the
`$\rightarrow$' direction. For example, the $r_i$ used in
(\ref{eq:addequ4}) are NOT those obtained from (\ref{eq:addequ3}),
but those obtained from the prior iteration for a message from $x'$
to $a$.

We note from the \emph{virtual} Tanner Graph in Fig.
\ref{fig:TannerCCRESM} that there is no add node in the top row.
Thus, the message going into the node $x[1]$ is simply set to
$(p_0,p_1)$ (note: $p_2=0$ always).

\subsubsection{From code nodes to check nodes}\label{32}
With reference to Fig. \ref{fig:CodeToCheck}(b-1), we now derive the
update equations for the message $U^ \leftarrow   = (u_0 ,u_1 )$
from a code node to a check node. As shown in the figure, the three
input messages are $Q^ \leftarrow   = (q_0 ,q_1 )$, $R^ \leftarrow =
(r_0 ,r_1 )$ and $W^ \to   = (w_0 ,w_1 )$. Based on the compatible
sum-product principle of the BP algorithm, we have
\begin{equation}
u_0  = \beta _{cu} q_0 r_0 w_0,\label{eq:varcequ1}
\end{equation}
where $\beta _{cu}$ is a normalization factor to ensure that the sum
of probabilities, $u_0  + u_1  = 1$. Similarly,
\begin{equation}
u_1  = \beta _{cu} q_1 r_1 w_1.\label{eq:varcequ2}
\end{equation}
So the overall message update associate with the compatibility
requirement of a code node is given by
\begin{equation}
U^ \leftarrow   = VAR_C (Q^ \to  ,R^ \to  ,W^ \leftarrow  ) = \beta
_{cu} (q_0 r_0 w_0 ,q_1 r_1 w_1 ).\label{eq:varcequ3}
\end{equation}
Similarly, with reference to Fig. \ref{fig:CodeToCheck}(b-2), we can
obtain the update equation for message $W^\leftarrow$:
\begin{equation}
W^ \leftarrow   = VAR_C (Q^ \leftarrow  ,R^ \leftarrow  ,U^ \to  ) =
\beta _{cw} (q_0 r_0 u_0 ,q_1 r_1 u_1 ).\label{eq:varcequ4}
\end{equation}

\subsubsection{From check nodes to source nodes}\label{33}
\begin{figure}
   \begin{minipage}[t]{0.49\linewidth}
     \centering
      \includegraphics[width=1\textwidth]{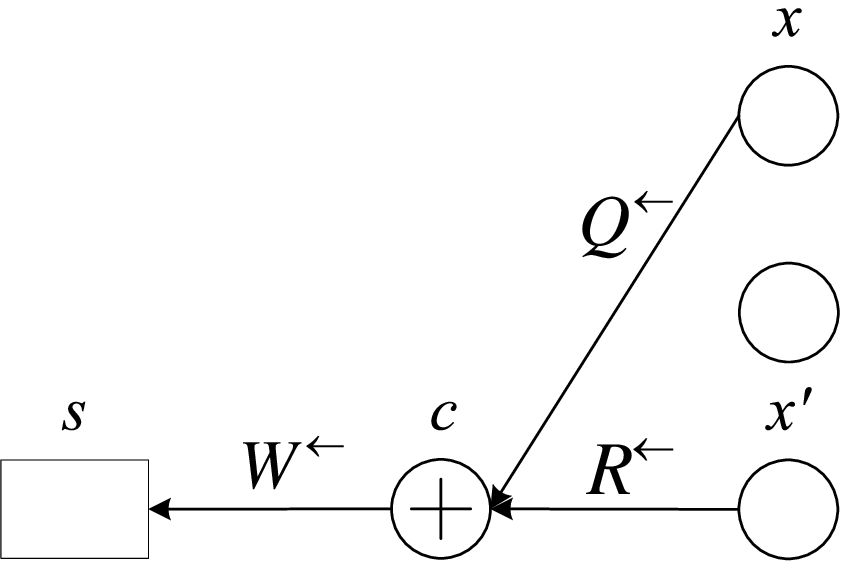}
      \caption{Update of message from check node to source node.} \label{Fig:CheckToSource}
   \end{minipage}
   \begin{minipage}[t]{0.49\linewidth}
      \centering
      \includegraphics[width=0.65\textwidth]{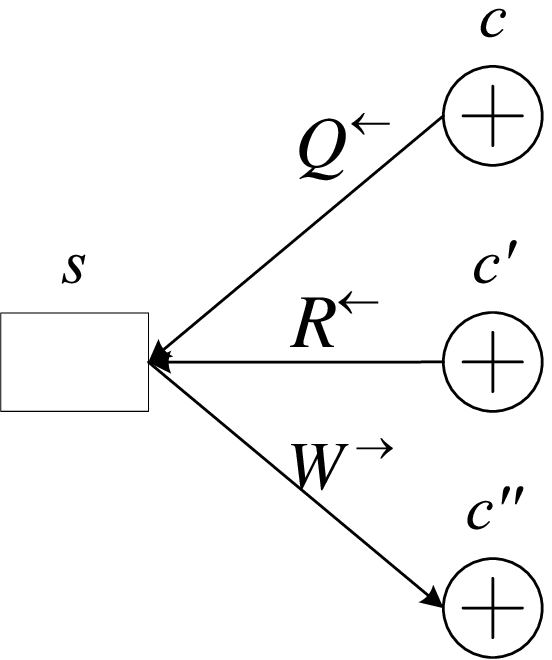}
      \caption{Update of message from source node to check node.} \label{Fig:SourceToCheck}
   \end{minipage}
\end{figure}

With reference to Fig. \ref{Fig:CheckToSource}, we now derive the
update equations for the message $W^ \leftarrow   = (w_0 ,w_1 )$
from a check node to a source node. The symbol \emph{s} can only be
0 if $(x,x')=(0,0)$ or $(x,x')=(1,1)$ (\emph{i.e.}, for
compatibility, $x \oplus x'$ must be \emph{s}). Based on the
sum-product principle of the BP algorithm, we have
\begin{equation}
w_0  = q_0 r_0  + q_1 r_1.\label{eq:chkequ1}
\end{equation}
Correspondingly $w_1$ can be obtained in a similar way. So we have
the messages going out from the check node:
\begin{equation}
W^ \leftarrow   = CHK(Q^ \leftarrow  ,R^ \leftarrow  ) = (q_0 r_0 +
q_1 r_1 ,q_0 r_1  + q_1 r_0 ).\label{eq:chkequ2}
\end{equation}

\subsubsection{From source nodes to check nodes}\label{34}
We now proceed to the updates of messages flowing from left to
right. With reference to Fig. \ref{Fig:SourceToCheck}, we derive the
message update rules for the message $W^ \to   = (w_0 ,w_1 )$ from a
source node to a check node. Again, based on the compatible
sum-product principle of BP, we have
\begin{equation}
(w_0 ,w_1 ) = (\beta _{sw} q_0 r_0 ,\beta _{sw} q_1 r_1 ),
\label{eq:varsequ1}
\end{equation}
where $\beta _{sw}$ is a normalization factor to ensure $w_0+w_1=1$.
So we have the output message of the source node as follows:
\begin{equation}
W^ \to   = VAR_S (Q^ \leftarrow  ,R^ \leftarrow  ) = \beta _{sw}
(q_0 r_0 ,q_1 r_1 ). \label{eq:varsequ2}
\end{equation}
Similarly we have
\begin{equation}
Q^ \to   = VAR_S (R^ \leftarrow  ,W^ \leftarrow  ) = \beta _{sq}
(w_0 r_0 ,w_1 r_1 ). \label{eq:varsequ3}
\end{equation}
\begin{equation}
R^ \to   = VAR_S (Q^ \leftarrow  ,W^ \leftarrow  ) = \beta _{sr}
(q_0 w_0 ,q_1 w_1 ). \label{eq:varsequ4}
\end{equation}

\subsubsection{From check nodes to code nodes}\label{35}
The update equations for the message from a check node to a code
node are similar to those from a check to a source node in Part
\emph{3)}, except for the update direction. So we omit the details
here.

\subsubsection{From code nodes to add nodes}\label{36}
The update equations for the message from a code node to an add node
are similar to those from a code to a check node in Part \emph{2)},
except for the update direction. So we also omit the details
here.$\\$

After the above parts are iterated and the values of the messages
converge, we declare that $s[k] = \arg \max _i (q_i r_i w_i )$,
where $q_i $, $r_i $, and $w_i $ are obtained from the three
messages $Q^ \leftarrow $, $R^ \leftarrow $ and $W^ \leftarrow $
flowing into the source node.

\section{Comparison of Different Methods}

We consider three channel-coded collisions resolution methods,
classified by how symbol misalignment is exploited in the
channel-decoding process. The first method performs the channel
decoding after the application of CRESM or A-MUD on the collided
signal. Thus, the channel decoding and collision resolution process
are decoupled. The second method is to apply Turbo-SIC to
iteratively decode the collided packets by applying a BP algorithm,
which is done in packet level. The third method is C-CRESM. As
presented in the preceding section, C-CRESM jointly decodes the
channel-coded misaligned signals at the symbol level.

\begin{figure}[tt]
\centering
\includegraphics[width=0.35\textwidth]{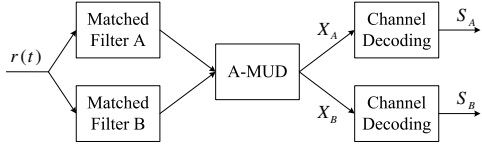}
\caption{Flowchart of Independent MUD-CD with two users.}
\label{fig:IndepMUDCD}
\end{figure}

\subsection{Independent Multiuser Detection and Channel Decoding (Independent MU-CD)}
The structure of the decoding process is shown in Fig.
\ref{fig:IndepMUDCD}. First the receive signals are passed through
two matched filters; then A-MUD is applied on the MF outputs to get
the channel coded packets; finally the channel coded symbols are
decoded by standard channel decoding. It could be understood
intuitively that separating the collision resolution and
channel-decoding processes may not be optimal. The channel-coded
bits are dependent on each other, but CRESM or A-MUD does not
exploit this property in the collision resolution process. This
method, however, is simple.

\begin{figure}[tt]
\centering
\includegraphics[width=0.35\textwidth]{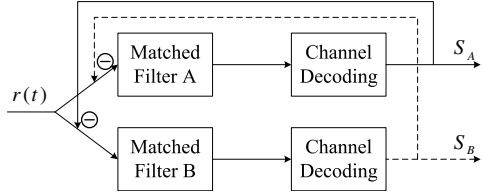}
\caption{Flowchart of Turbo-SIC with two users.}
\label{fig:TurboSIC}
\end{figure}

\begin{figure}[tt]
\centering
\includegraphics[width=0.45\textwidth]{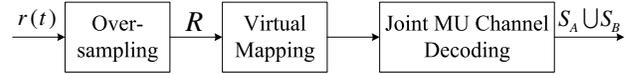}
\caption{Flowchart of C-CRESM with two users.} \label{fig:CCRESM}
\end{figure}

\subsection{Turbo-SIC}
Turbo-SIC was proposed in \cite{WangPoorJASC99} to decode
channel-coded packets in a CDMA system. The structure of the
decoding process is shown in Fig. \ref{fig:TurboSIC}. Turbo-SIC
digitalizes the signals by matched filtering, and thus relies
heavily on the crosscorrelation of signature waves. In the CDMA
system, the signature waves of different users are different
(\emph{i.e.}, the matched filters A and B in Fig. \ref{fig:TurboSIC}
correspond to near-orthogonal signature waves). In the system of
interest to us here, the signature waves are the same except for the
symbol misalignment. The system will certainly not perform well when
the symbol misalignment is zero. For non-zero symbol misalignment,
the performance will improve.

\begin{figure*}
   \begin{minipage}[t]{0.33\linewidth}
     \centering
      \includegraphics[width=1.1\textwidth]{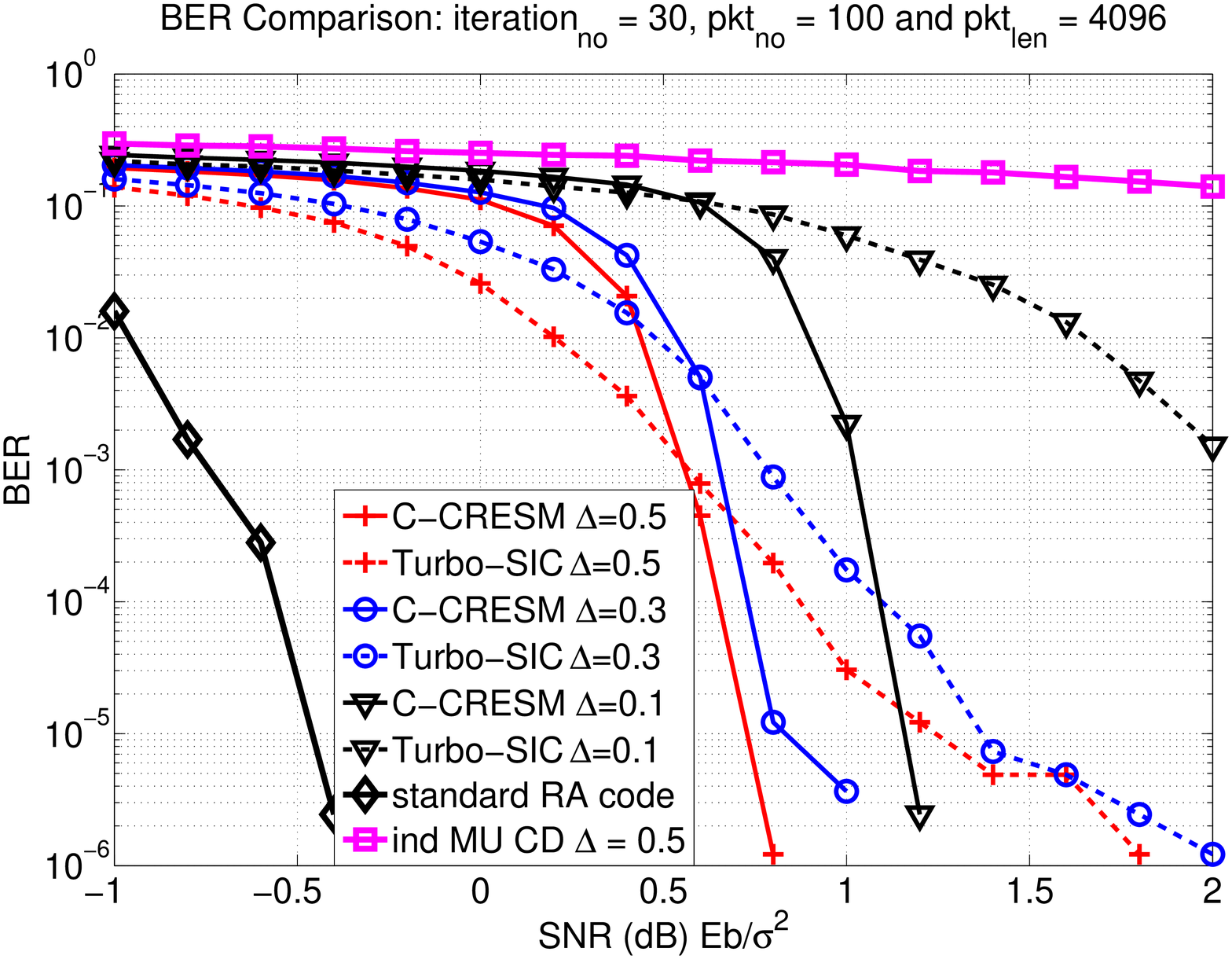}
      \caption{BER Comparison of C-CRESM and Turbo-SIC for different $\Delta$.}
      \label{fig:BERcomparison1}
   \end{minipage}
   \begin{minipage}[t]{0.33\linewidth}
      \centering
      \includegraphics[width=1.1\textwidth]{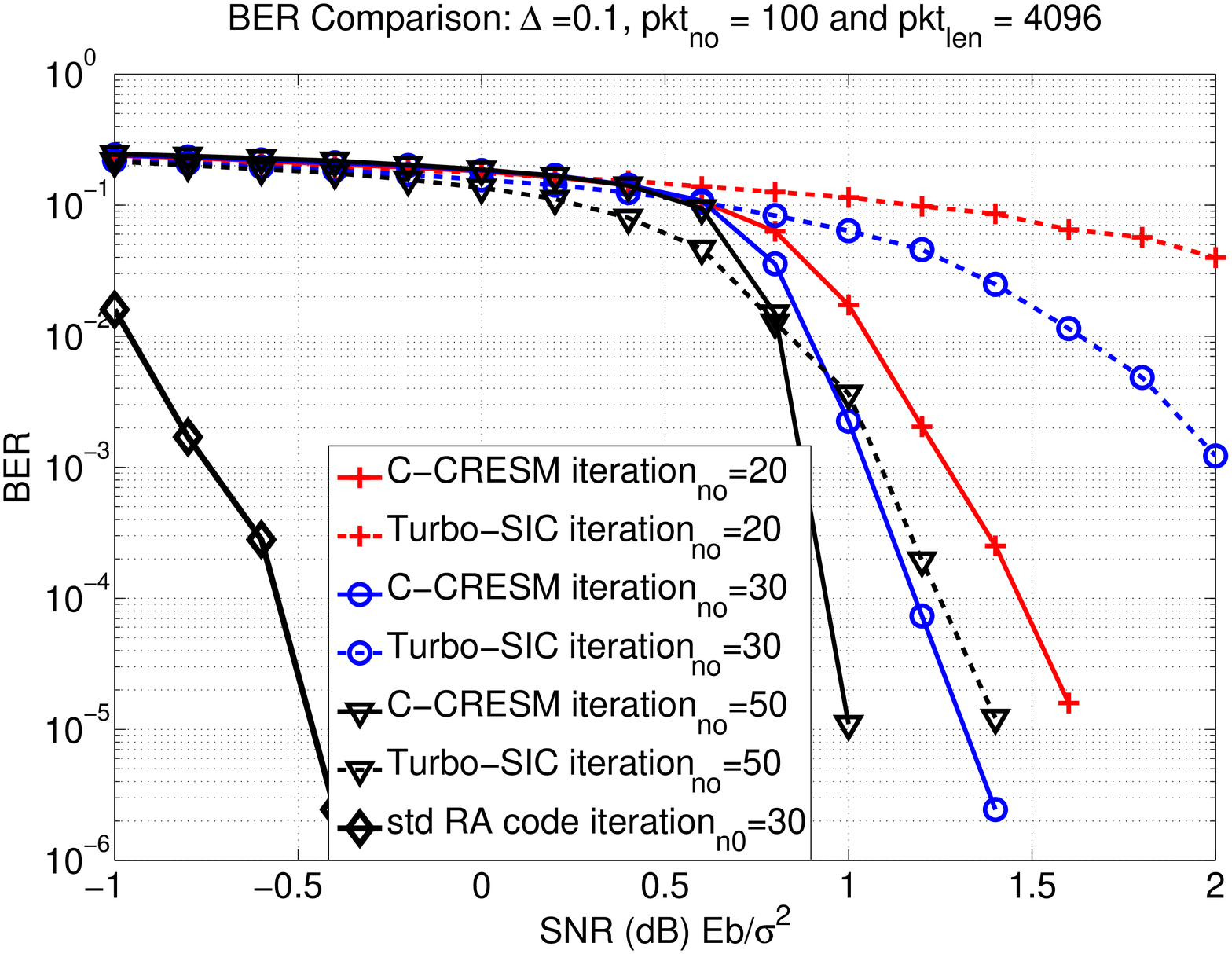}
      \caption{BER Comparison of C-CRESM and Turbo-SIC for different iteration numbers.}
      \label{fig:BERcomparison2}
   \end{minipage}
   \begin{minipage}[t]{0.33\linewidth}
     \centering
      \includegraphics[width=1.1\textwidth]{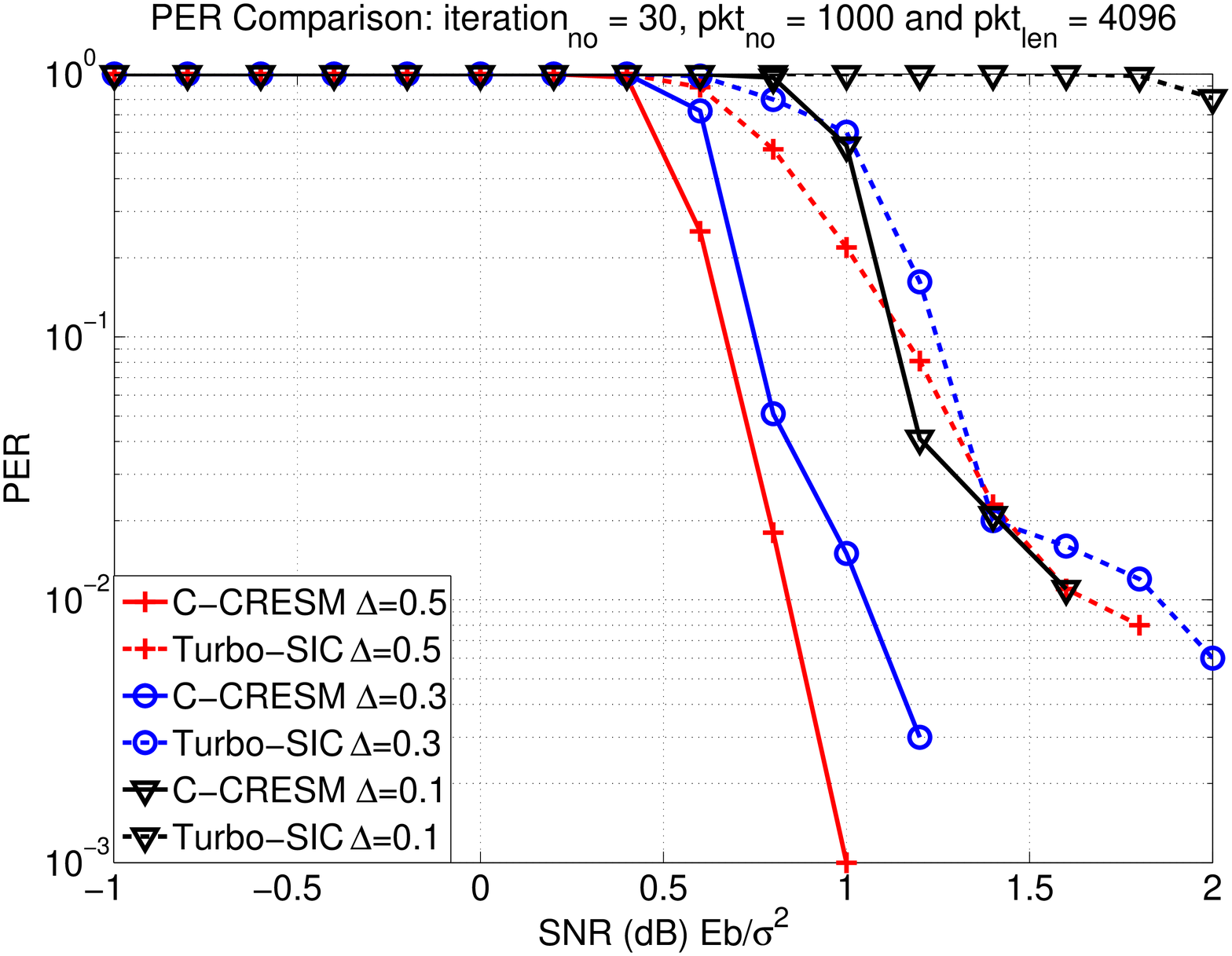}
      \caption{PER Comparison of C-CRESM and Turbo-SIC for different $\Delta$.}
      \label{fig:PERcomparison1}
   \end{minipage}
\end{figure*}


With reference to Fig. \ref{fig:TurboSIC} again, the structure of
the decoding process is under the framework of successive
interference cancellation (SIC) \cite{SICMobicom08 , TseBook}. The
channel decoding is performed for each signal separately after
matched filtering. The soft information of the separately decoded
signals are then fed back to the front end for interference
cancellation purposes. The process of matched filtering, channel
decoding, and interference cancellation is iterated to reduce the
error probability progressively. Within each iteration, the symbols
from different branches of the decoder do not ``interact''. So, this
method is a packet-level joint MUD channel decoding.

\subsection{Channel-coded CRESM (C-CRESM)}
Symbol misalignment of collided signals result in a ``virtual
signal'' with ``virtual symbols'' at twice the rate of the original
symbols from each source, after being oversampled. These virtual
symbols are correlated in two ways. First, adjacent virtual symbols
are correlated via misalignment of original symbols. Second, they
are correlated via the channel coding. The idea of C-CRESM is to
make use of these correlations to better decode the underlying
symbols embedded in the virtual symbols. In this paper, we assume
the use of RA channel code. C-CRESM maps the virtual symbols to
evidence nodes in the virtual Tanner graph shown in Fig.
\ref{fig:CCRESM} and Fig. \ref{fig:TannerCCRESM}. It then uses BP to
decode the underlying symbols.

\section{Simulation Results}
In this section we present simulation results to illustrate the
performance of C-CRESM compared with Turbo-SIC (the Turbo-SIC method
outperforms the traditional separate channel decoding and MUD
method, which is represented by the magenta line with square markers
on it in Fig. \ref{fig:BERcomparison1}.). In all the simulations, we
assume BPSK modulation with phase synchronization. The repeat factor
for the RA code is $q=3$ with an interleaver that is randomly
selected for each packet but identical for C-CRESM and Turbo-SIC.
Note that here we only consider the average results of different
interleavers, but in practice we should fix the system to a good
interleaver. The choice of an interleaver is beyond the scope of
this paper, however. We refer the reader to \cite{RAcodeJinPhd} for
details.

We assume the noise is AWGN with variance $\sigma^2$ and define the
SNR as $\frac{{E_b }}{{\sigma ^2 /2}} = \frac{{E_A  + E_B }}{{\sigma
^2 /2}} = \frac{1}{{\sigma ^2 }}$ (\emph{i.e.}, the signals from
both ends have unit power.). For our simulations, we modify
Turbo-SIC in \cite{WangPoorJASC99} to also use the RA code for fair
comparison with C-CRESM. The Turbo iteration number (the outer loop
in Fig. \ref{fig:TurboSIC}) is $m$ and the RA decoding iteration
number is $n$ (the iteration within the channel decoding block in
Fig. \ref{fig:TurboSIC}). Thus, the total number of iterations in
Turbo-SIC is $mn$. We also set the iteration number in C-CRESM to be
$mn$ for fair comparison. The computation costs within each
iteration of the schemes are the same, which can be easily derived
from the message update rules and message set.

The black line with a `diamond' marker in Fig.
\ref{fig:BERcomparison1} and Fig. \ref{fig:BERcomparison2} is the
standard one to one RA code communication result, served as a
benchmark for comparison. Fig. \ref{fig:BERcomparison1} shows for
the target BER below $10^{-3}$, C-CRESM performs better than
Turbo-SIC by 0.25 to 1.5 dB for different $\Delta$. Fig.
\ref{fig:BERcomparison2} compares C-CRESM with Turbo-SIC for
different iteration numbers, fixing $\Delta = 0.1$. For both
schemes, the performance improves as iteration number increases, and
C-CRESM has a 0.3 to 1 dB improvement at the BER of $10^{-3}$.  Both
figures reveal that as SNR increases, the SNR-BER curve of C-CRESM
drops faster than that of Turbo-SIC. Fig. \ref{fig:PERcomparison1}
presents the packet error rate (PER) for various $\Delta$. Again,
C-CRESM performs better than Turbo-SIC by at least 0.75 dB for
target PER below $10^{-2}$. For both BER and PER, the advantage of
C-CRESM over Turbo-SIC is particularly striking when $\Delta = 0.1$.
Since C-CRESM can work effectively over a wider range or $\Delta$,
it is robust when deployed in the field, where there is no
deliberate synchronization to control $\Delta$.

\section{Conclusion}
This paper has proposed and investigated C-CRESM, a novel
channel-coded collision resolution method, targeting at 802.11-like
WLANs. This method integrates the collision resolution process and
the channel decoding process into one mechanism. A key ingredient in
this method is the observation that symbols of collided signals in a
WLAN are most likely not aligned when they arrive at the receiver.
This is because there is no deliberate attempt to synchronize the
transmitters at the symbol level. It turns out that such symbol
misalignment is good because it introduces a certain degree of
orthogonality between the otherwise similar signal waveforms of the
collided signals.

C-CRESM exploits the symbol misalignment in the channel decoding
process. In particular, we show that when the RA channel code is
used, the collided signals at the receiver (from multiple sources)
can be thought of as a virtual signal (from single source). The
virtual signal corresponds to  a virtual RA channel code, which is a
superposition of the overlapped, misaligned, RA codewords in a
collision. This concept allows us to construct a {\it virtual}
Tanner graph for an integrated
collision-resolution-and-channel-decoding process.

The integrated decoding process in C-CRESM operates at the symbol
level. Our simulation results show that it outperforms the
straightforward method in which collision resolution and channel
decoding are performed in an independent way, as well as the method
in which collision resolution and channel decoding are integrated at
the packet level (Turbo-SIC). Compared with independent collision
resolution and channel decoding, C-CRESM have a more than 3dB gain.
Compared with Turbo-SIC, C-CRESM has more than 0.75dB gain when the
target PER is $10^{-2}$ for typical packet lengths. C-CRESM is also
more robust against the variation of the symbol misalignment
compared with these other methods.



%
%

\end{document}